\documentclass[noshowpacs,amsmath,twocolumn,superscriptaddress,10pt,aps,prb]{revtex4-1} 
\usepackage{setspace}
\usepackage{amsmath}
\usepackage{breqn}
\usepackage{graphicx}
\usepackage[nearskip,margin = 0pt,caption=false]{subfig}

\usepackage{verbatim}
\usepackage{amsfonts}
\usepackage{amssymb}
\usepackage{epstopdf}
\usepackage{lipsum}
\usepackage{siunitx}
\usepackage{xcolor}
\usepackage{array}
\usepackage[normalem]{ulem}
\usepackage[resetlabels,labeled]{multibib}
\DeclareGraphicsExtensions{.pdf,.eps,.png,.jpg,.mps}
\usepackage{etoolbox}

\begin{document}
\title{Inverse-designed Silicon Carbide Quantum and Nonlinear Photonics}
\author{Joshua Yang$^{1}$, Melissa A. Guidry$^{1}$, Daniil M. Lukin$^{1}$, Kiyoul Yang$^{1,2}$, and Jelena Vu\v{c}kovi\'{c}$^{1,*}$\\
\vspace{+0.05 in}
$^1$E.L.Ginzton Laboratory, Stanford University, Stanford, CA, USA.\\
$^2$John A. Paulson School of Engineering and Applied Sciences, Harvard University, Cambridge, MA, USA.\\
{\small *Corresponding author: jela@stanford.edu}}

\begin{abstract}
\noindent 
\color{black} Inverse design has revolutionized the field of photonics, enabling automated development of complex structures and geometries with unique functionalities unmatched by classical design. However, the use of inverse design in nonlinear photonics has been limited. In this work, we demonstrate quantum and classical nonlinear light generation in silicon carbide nanophotonic inverse-designed Fabry-Pérot cavities. We achieve ultra-low reflector losses while targeting a pre-specified anomalous dispersion to reach optical parametric oscillation. By controlling dispersion through inverse design, we target a second-order phase-matching condition to realize second- and third-order nonlinear light generation in our devices, thereby extending stimulated parametric processes into the visible spectrum. This first realization of computational optimization for nonlinear light generation highlights the power of inverse design for nonlinear optics, in particular when combined with highly nonlinear materials such as silicon carbide.
\end{abstract}

\maketitle
 
\begin{figure*}[t!]
\centering
\includegraphics[width=0.8\linewidth]{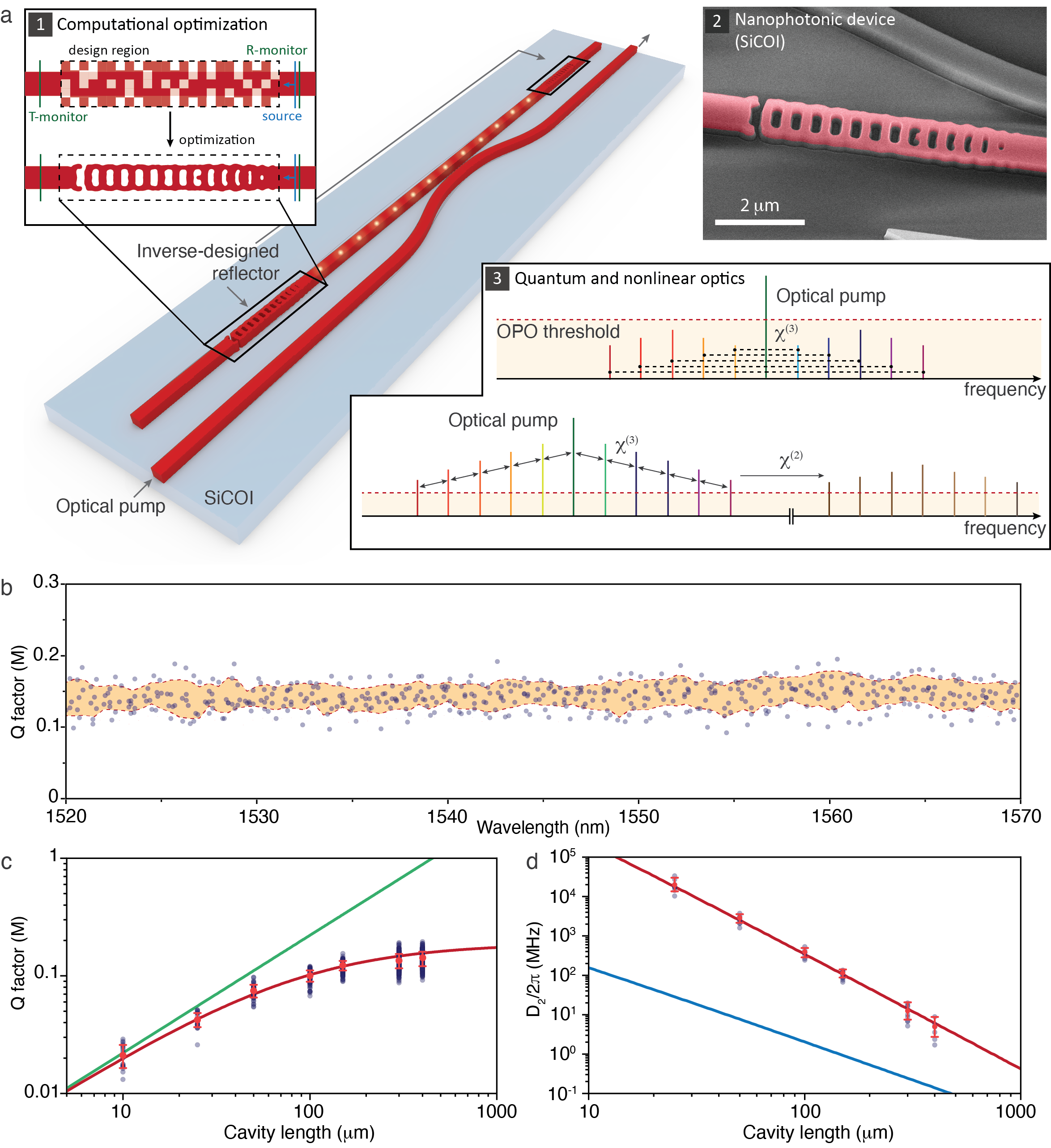}
\caption{\label{fig:Fig1}{\bf{An inverse-designed silicon carbide optical cavity}} (\textbf{a}) Schematic of the device. Light is delivered throughout the bus waveguide which couples evanescently to the waveguide section of the resonator. Inset 1: Computational design of on-chip reflector structure using gradient-based optimization. Inset 2: A SEM image of an inverse-designed reflector (false red color) prior to the SiO$_{2}$ encapsulation. Inset 3: Illustration of the nonlinear processes studied in this work: the $\chi^{(3)}$ nonlinearity creates broadband parametric gain, resulting in comb formation. The generated comb is then translated into visible light via sum frequency generation through the $\chi^{(2)}$ nonlinearity. (\textbf{b}) Quality factors of resonances for 300 $\mu$m- and 400 $\mu$m- long devices plotted against resonance wavelength, with regions in beige indicating data within a standard deviation of the mean, showing robust performance. (\textbf{c}) Scatter plot of intrinsic quality factors for varying cavity lengths, including fitted quality factor curves for FP devices with reflectivites of 99.56$\%$ with (red) and without (green) waveguide loss of 3.0 dB/cm. (\textbf{d}) Plot of measured dispersion versus cavity length. Fitted curve (red) with a dispersion relationship proportional to L$^{-3}$, dominated by the dispersion of the inverse-designed reflector. Simulated curve (blue) for the waveguide-only dispersion, proportional to L$^{-2}$.  }
\end{figure*}

\section{Introduction}

\color{black} 

To further advance capabilities of chip-scale quantum and nonlinear photonics, inverse design using gradient-based optimization\cite{Piggott:2015:NaturePhotonics,Su:2020:APR} has attracted considerable attention to realize novel functionalities along with operational robustness and compact device size\cite{Piggott:2020:ACSPhotonics}. Primarily in silicon photonics\cite{Piggott:2020:ACSPhotonics}, there have been numerous experimental demonstrations of inverse-designed devices as well as system applications, such as particle accelerators\cite{Sapra:2020:Science}, optical ranging\cite{Yang:2020:NaturePhotonics}, and communications\cite{Yang:2022:NatureComm}. Beyond silicon, inverse design has also been applied to other photonic platforms, including diamond\cite{Dory:2019:NatureComm}, silicon carbide\cite{Guidry:2020:Optica}, lithium niobate\cite{Shang:LN:2022:arXiv}, and materials like chalcogenide glasses\cite{Lin:Chalcogenide:2022:LPR}.
Nonetheless, inverse design for nonlinear photonics has been limited. To efficiently trigger optical nonlinearities at quantum or classical optics regimes, extremely low scattering losses are required, despite the highly irregular geometries in inverse-designed structures that do not offer intrinsic symmetry-based protection against such losses. While low-loss inverse-designed reflectors have recently been demonstrated in silicon \cite{Ahn:2022:ACSPhotonics}, nonlinear photonics also requires a material that features suitable nonlinear optical characteristics while simultaneously possessing favorable fabrication properties.

Silicon carbide (SiC) has emerged as a promising material platform for both quantum and nonlinear photonics. In addition to hosting optically addressable color centers for integrated quantum technologies\cite{Awschalom:2018:NaturePhotonics,Wrachtrup:2022:NatureMaterials,Awschalom:2020:NanoLetters}, SiC exhibits compelling optical properties, including a high refractive index, large second- and third-order nonlinearities, and a broad optical transparency window\cite{Lin:2014:OE,Lipson:2015:OL}. The combination of all these strengths with established SiC manufacturing technologies in the electronics industry has positioned the SiC platform as a promising, CMOS-compatible candidate for the development of next-generation monolithic photonic-electronic systems. Developments in SiC-on-insulator (SiCOI) production\cite{Lukin:2020:NaturePhotonics, Noda:2019:Optica, Fan:18} have accelerated recent developments in optoelectronics\cite{Loncar:2022:NatureCommun}, quantum\cite{Lukin:2020:NaturePhotonics,Lukin:2022:arXiv} and nonlinear photonics\cite{Noda:2019:Optica,Lukin:2020:NaturePhotonics,Guidry:2020:Optica,Guidry:2022:NaturePhotonics,Wang:2021:Light,Cai:2022:PhotonResearch}.

In this work, we leverage the combination of powerful inverse design techniques with favorable nonlinear optical properties of SiC, realizing optical Fabry-Pérot (FP) cavities in SiCOI using inverse-designed reflectors for quantum and nonlinear photonic experiments. The inverse-designed cavity is optimized for low scattering loss and fabrication error robustness over multiple wavelengths while simultaneously constrained for a pre-specified dispersion; the final reflector structure occupies a compact footprint of 6.75 $\mu$m by 1 $\mu$m. Using the inverse-designed optical cavities, we present a spontaneous four-wave mixing, resulting in signal and idler photon pair generations around a telecom pump, with a coincidence to accidental ratio (CAR) of 275 at a photon generation rate of 0.1 MHz. In addition, we demonstrate stimulated parametric oscillations in the C-band and, simultaneously, nonlinear frequency generation at visible wavelengths, via third- and second-order nonlinear processes, respectively. We show robustness of the device performance across the optimized reflectivity band, producible across the chip, with reflectivity limited by waveguide losses rather than the inverse-designed reflector, showing clear avenues for significant improvements in future devices. Our results showcase the utility of inverse-designed FP cavities, and to our knowledge, are the first experimental demonstrations of inverse design in quantum and nonlinear light generation, serving as a key addition to the general toolbox of quantum and nonlinear photonics.

\begin{figure*}[t!]
\centering
\includegraphics[width=0.8\linewidth]{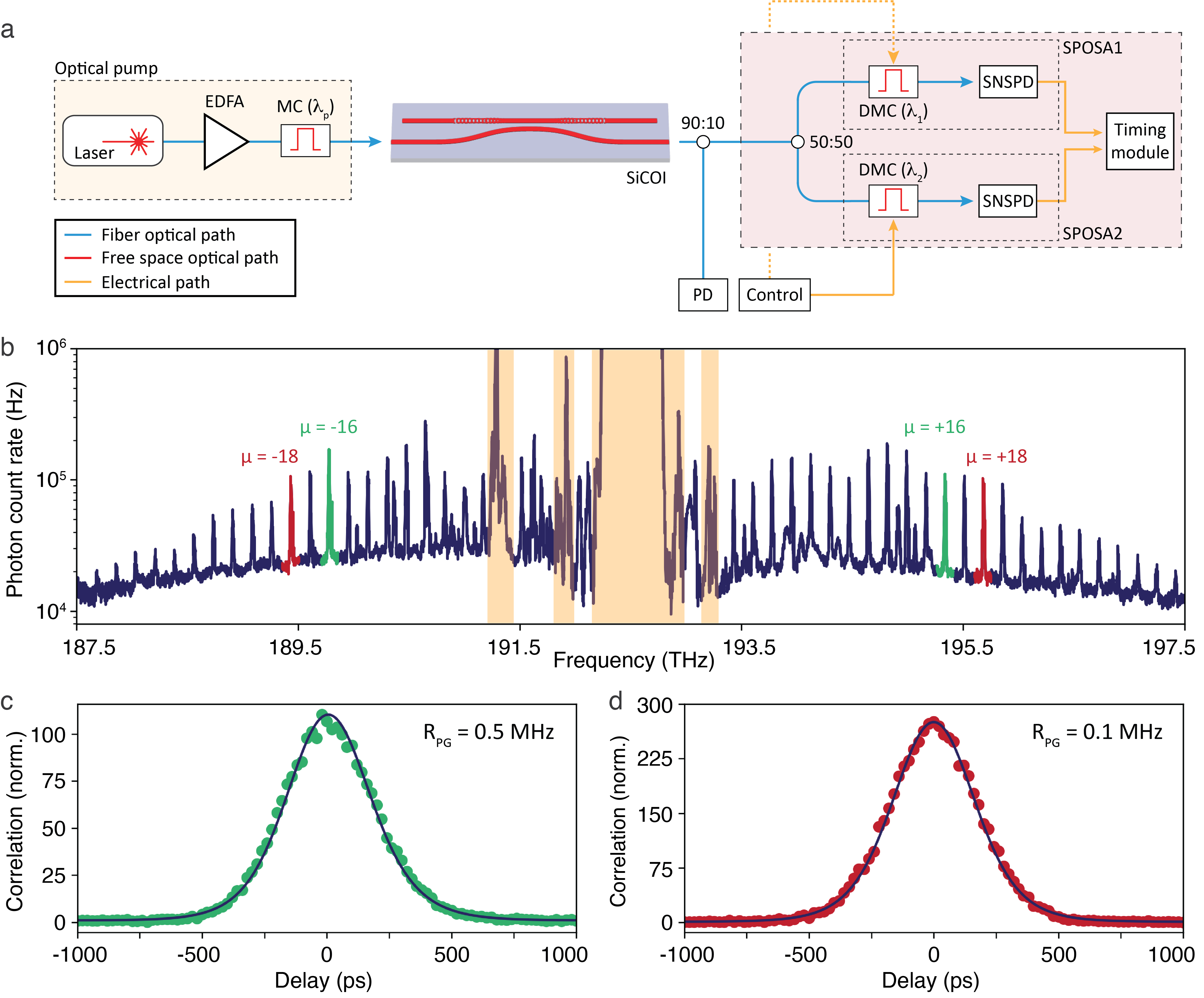}
\caption{\label{fig:Fig2}{\bf{Below-threshold quantum comb}} 
(\textbf{a}) 
Diagram of the measurement setup used in the quantum comb generation experiments (EDFA: Erbium-doped fiber amplifier, MC: Monochromator, DMC: Double monochromator, PD: Photodetector, SNSPD: Superconducting nanowire single-photon detector)
(\textbf{b}) Below-threshold Kerr frequency comb formation observed on the single photon optical spectrum analyzer (SPOSA). Regions highlighted in beige indicate frequencies with high pump noise.
(\textbf{c,d}) Cross-correlation measurements taken on mode numbers $\mu = \pm 16$ and $\mu = \pm 18$, with CAR ratios of 110 and 275, at pair generation rates of 0.5 MHz and 0.1 MHz, respectively.}
\end{figure*}

\begin{figure*}[t!]
\centering
\includegraphics[width=0.80\linewidth]{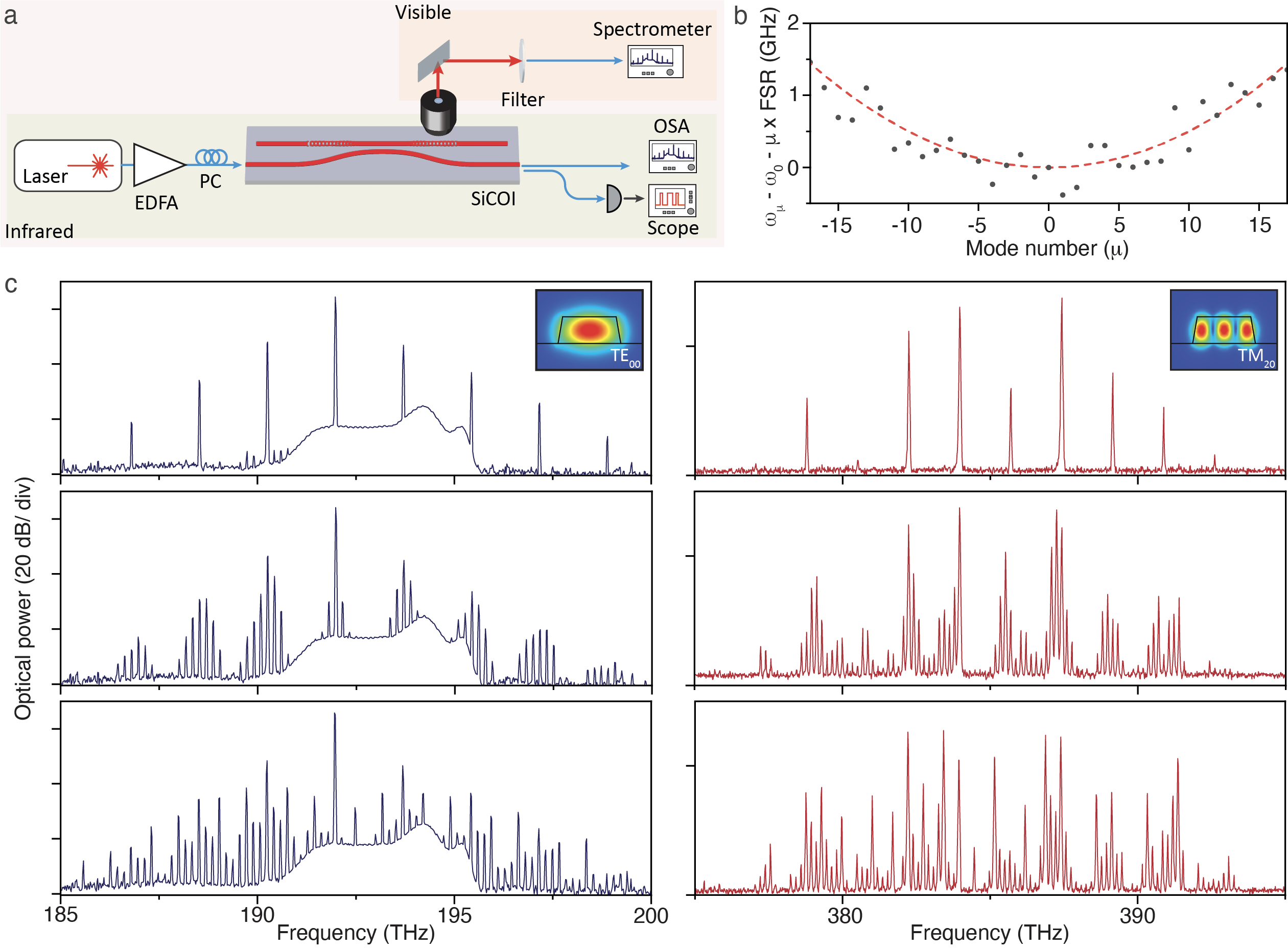}
\caption{\label{fig:Fig3}{\bf{Microcomb formation in an inverse-designed silicon carbide resonator}} 
(\textbf{a}) Diagram of the measurement setup used in the nonlinear light generation experiments (EDFA: Erbium-doped fiber amplifier, PC: polarization controller, OSA: optical spectrum analyzer). 
(\textbf{b}) Measured integrated dispersion of the FP device with respect to relative mode number $\mu$, with the center mode at $\mu = 0$ (frequency = 194 THz). The dotted line plots a numerical fit of the dispersion, with parameters $D_1 = 174.8$ GHz and $D_2/2 \pi = 10.1$ MHz. 
(\textbf{c}) Measured OPO telecom spectra and SFG visible spectra at different stages of microcomb formation, with a threshold power of 400 mW. Insets: simulated profiles of the phase-matched telecom and visible modes. }
\end{figure*} 

\section{Results}

\subsection{Device design and implementation}

Optical FP cavities were deployed due to several considerations. Firstly, analogous to table-top ultrafast laser cavities, chip-scale FP structures offer more design flexibility\cite{Yu:NIST:2019:ACSPhotonics,Ahn:2022:ACSPhotonics,Herr:2022:arXiv} compared to whispering-gallery-mode (WGM) cavities\cite{Gaeta:2014:OL,Vahala:2016:NaturePhotonics,Kippenberg:2016:Science}. Additionally, FP structures can support multiple transverse modes over a broad spectrum and are not limited by other physical rules in terms of the frequency spacing between the transverse modes, which is important for optimizing quantum and nonlinear optical processes on a chip. This is in contrast to photonic crystal and WGM cavities\cite{DeRossi:2021:NaturePhotonics}. Moreover, in comparison to microring resonators, FP structures offer an additional degree of freedom in device design, allowing for dispersion dominated by reflector design and not limited by the effective waveguide mode of the cavity.   

Using photonic inverse design, reflector structures interfaced to a standard, single-mode rectangular waveguide were designed to form a high-Q, dispersion-engineered FP cavity (see Fig.\ref{fig:Fig1}a and Inset 1); intensity and phase of the reflected mode determine Q-factor and group velocity dispersion of the cavity, respectively\cite{Herr:2022:arXiv,Ahn:2022:ACSPhotonics}. The design process of the reflector structures also utilizes an extra broadband performance optimization constraint by simultaneously optimizing reflectivities at multiple wavelengths centered around the target wavelength \cite{sapra2019inverse}. The optimized structures are fabricated in thin-film 4H-SiCOI\cite{Lukin:2020:NaturePhotonics} made from high-purity semi-insulation (HPSI) 4H-SiC wafers from Cree, which features a high Kerr nonlinear refractive index ($n_2 = 9.1 \pm 1.2$)\cite{qingli}. The devices are patterned through an electron-beam lithographic process detailed in Ref. \cite{Guidry:2020:Optica}. It is noteworthy that all the device structures comply with the process design rules for standard commercial foundries.  

For device characterizations, light is coupled on and off the chip to a bus waveguide either confocally with an objective lens through compact inverse-designed grating couplers\cite{Guidry:2020:Optica} or through the chip facet with a lensed fiber. The bus waveguide is side coupled, as shown in Fig.\ref{fig:Fig1}a to the FP cavity waveguide. Resonator Q-factors and dispersion of the FP devices are simultaneously measured by sweeping a continuously tunable mode-hop-free laser from 1520 nm to 1570 nm. Q-factors were extracted through a Fano fit of the resonant transmission dips, and the dispersion of each device was measured by utilizing a calibrated Mach-Zehnder fiber interferometer\cite{Vahala:2018:NaturePhotonics}. This processing then allows us to characterize the repeatability of the device performance across the optimized telecom reflectivity band. We examine 16 devices with cavity lengths (defined as the length of the straight waveguide between the reflectors) of 300 $\mu$m and 400 $\mu$m, to satisfy the requirements of high Q-factor and anomalous dispersion. Their Q-factors across the full range of the C-band tunable laser are shown in Fig.\ref{fig:Fig1}b, revealing robust and broadband performance across the telecom band, enabled by the aforementioned inverse-design optimization constraints.

To further understand and isolate the characteristics of loss and dispersion of the waveguide from that of the reflectors, additional FP devices with cavity lengths ranging from 10 to 400 $\mu$m were examined, and the resulting Q-factors and dispersion of the resonators are plotted in Fig.\ref{fig:Fig1}c,d. As shown in Fig.\ref{fig:Fig1}c, the Q-factors saturate with increasing cavity length as the waveguide transmission loss dominates over the optical losses of the reflectors\cite{Yu:NIST:2019:ACSPhotonics}. We extract a waveguide loss of 3.0 dB/cm, consistent with the performance expected for high-confinement single-mode waveguides, which are more susceptible to scattering losses due to sidewall roughness as well as fabrication imperfections\cite{Lipson:2017:Optica}. However, this dimension is necessary for second-harmonic generation through phase-matching. Upon accounting for this waveguide loss, the analysis shows broadband average reflectivity of approximately 99.56$\%$ in the inverse-designed reflectors, as obtained from the fit across 65 FP devices with 1069 resonances and seven different cavity lengths (Fig.\ref{fig:Fig1}c).

In Fig.\ref{fig:Fig1}d, dispersion of the FP resonators are plotted for devices that have at least three resonances within the tunable laser range (1520 - 1570 nm). Dispersion of the highly anomalous cavities was plotted alongside the simulated dispersion from the waveguides. The dispersion resulting from the waveguide geometry follows an expected inverse relationship with the cavity length, proportional to L$^{-2}$, while the measured cavity dispersion is proportional to L$^{-3}$. The L$^{-2}$ relationship is expected from the waveguide geometry, since $D_2 = \beta_2$L${D_1}^3$, where $D_1 \propto$ L$^{-1}$. Likewise, the L$^{-3}$ relationship from the cavity is also expected, as the dispersion of the FP cavity with the inverse-designed reflectors is orders of magnitude greater than that simulated from the waveguide geometry, suggesting that the inverse-designed reflectors are the dominant contributors to the dispersion of the device, which contributes the extra L$^{-1}$ term. 

By demonstrating that the dispersion of the FP devices are dominated by the reflectors rather than the waveguide geometry, the aforementioned limitations in Q-factors can be overcome in further optimizations while retaining single-mode behavior by designing fundamental-mode reflectors for wider multi-mode waveguides, which has been demonstrated previously in the silicon-on-insulator platform\cite{Ahn:2022:ACSPhotonics}. This would enable lower-power comb generation, but no longer satisfy the second-harmonic phase-matching condition.

\subsection{Cross-correlations of spontaneous four-wave mixing}

The use of inverse design for quantum photonics has several potential enabling applications, such as compact and efficient splitting and routing of signal-idler entangled photons on-chip and utilization in quantum circuits to allow for compactness and high performance. As an initial demonstration of the use of inverse design in nonlinear quantum light generation, below-threshold characteristics of a 300 $\mu$m long inverse-designed FP device were measured by using a single-photon optical spectrum analyzer (SPOSA)\cite{Guidry:2022:NaturePhotonics}. The SPOSA was constructed using a two-channel multipass grating monochromator setup, with the output sent to superconducting nanowire single-photon detectors (SNSPDs) from PhotonSpot. The erbium-doped fiber amplifier (EDFA) pump laser was double-filtered and tuned on resonance to the FP cavity, and the spectrum of the below-threshold quantum comb resulting from spontaneous parametric oscillations was collected (Fig.\ref{fig:Fig2}b). 

We then perform a cross-correlation measurement on two different signal-idler pairs using the two channels of the SPOSA and obtain a coincidence-to-accidental (CAR) ratio of 110 and 275, at photon pair generation rates of 0.5 MHz and 0.1 MHz, respectively. We fit the obtained measurements with a convolution of simulated cross-correlations using input-output theory \cite{1999PhRvL..83.2556O} and a gaussian response corresponding to a detector jitter of 200 ps, plotted in Fig.\ref{fig:Fig2}c,d. This demonstrates the potential for use of this inverse-designed FP device geometry as an entangled photon pair source, paving the way for complete integration of selective outcoupling, high brightness, compact pair sources in future inverse-designed devices.  

\subsection{Stimulated parametric oscillations}
In order to observe optical parametric oscillation (OPO), we select a 300 $\mu$m long FP cavity with suitably high Q-factors and anomalous dispersion, to satisfy the necessary frequency and phase-matching conditions. We send a tunable laser through an EDFA and tune the pump power onto the target FP cavity resonance. The output from the bus waveguide is sent to an optical spectrum analyzer (OSA). At a threshold power of 400 mW, we observe the formation of a primary comb spanning the C-band, shown in Figure 3, with primary sidebands appearing at mode numbers $\mu$ = $\pm$ 10, relative to the pump mode at 192 THz. As we continue to increase the pump power, we observe the formation of the secondary combs and the filling out of subcombs. 

As noted earlier, the waveguide section of the cavity achieves phase-matching between the fundamental mode TE$_{00}$ at 1550 nm and the transverse-magnetic mode TM$_{20}$ at 775 nm for second-order nonlinear processes, using the $d_{31}$ nonlinear term in 4H-SiC. As a result, even though the reflectors are not engineered as vertical couplers, they scatter out-of-plane a significant amount of visible light, generated during this four-wave-mixing process.
By outcoupling the scattered light through an objective and then sending it to a spectrometer, we observe a spectrally-translated frequency comb in the visible regime, a result of sum-frequency generation and second-harmonic generation in the FP cavity. While the inverse-designed reflectors were only optimized for high reflectivity at telecom, FDTD simulations also indicate moderate reflectivities for the TM$_{20}$ mode at visible wavelengths. This, in combination with the cavity waveguide geometry that satisfies the phase matching condition between the fundamental TE mode at telecom and the TM$_{20}$ mode at visible, results in the second-order nonlinear process that reproduces the shape of the C-band comb into the visible part of the spectrum.

\section{Discussion}

\noindent In conclusion, we have demonstrated the first experimental realization of inverse design in quantum and nonlinear light generation by utilizing optical inverse-designed FP cavities in 4H-SiCOI, with devices that support second- and third-order nonlinear processes.
We characterize stimulated parametric oscillation at telecom and visible frequencies. Limitations of performance on current devices due to waveguide loss can be overcome in future iterations by designing single-mode reflectors for wider multi-mode waveguide geometries, reducing scattering losses due to fabrication imperfections while retaining anomalous dispersive behaviour. Assuming a wider waveguide transmission loss as low as 0.08 dB/cm, demonstrated in previous SiC microring resonators\cite{Guidry:2022:NaturePhotonics} that correspond to a Q factor of 5$\times 10^{6}$, FP cavities with the inverse-designed reflectors would exhibit Q factors between 3$\times 10^{4}$ and 1.0$\times 10^{6}$ for cavity lengths in the range 10 - 500 $\mu$m.  
 By improving these Q factors of future devices with inverse-designed reflectors optimized for wider waveguides, we expect to lower the OPO threshold power of these devices by over an order of magnitude to below 20 mW. Furthermore, additional optimization of the inverse-designed reflectors at visible wavelengths would allow for more efficient second-order processes. Beyond classical photonics, we also demonstrate nonlinear light generation in the quantum regime, measuring photon pair generation and cross-correlations in below-threshold quantum combs. By displaying generation of quantum light with inverse design, we open the potential for future quantum light generation experiments, including, but not limited to, the use of numerical optimization for quantum filtering sections and selective signal-idler pair outcoupling for each reflector\cite{Piggott:2015:NaturePhotonics, doi:10.1021/acsphotonics.7b00987}. With inverse-designed devices, generation, filtering, and sorting of quantum light could be implemented all within a single architecture and optimized for foundry compatibility\cite{Piggott:2020:ACSPhotonics}. In total, these demonstrations showcase the utility and potential that inverse design brings to the fields of quantum and nonlinear photonics.

\section{Methods}

\noindent\textbf{Photonic inverse design} Stanford Photonics Inverse Design Software (SPINS) was used to optimize waveguide reflectors for target reflections of 100 \% for 1400, 1500, 1600, and 1700 nm fundamental TE-polarized waveguide modes. The design area is 6750 $\times$ 1000 nm$^{2}$, minimum feature size is 120 nm, and layer thickness is 350 nm. To engineer group velocity dispersion of FP cavities, phases of reflecting waves at different wavelengths were pre-determined as additional optimization targets.  \\

\noindent\textbf{Device Fabrication} 
Thin-films of SiCOI were produced from bulk, high-purity semi-insulation (HPSI) 4H-SiC Cree wafers that have been bonded to thermally oxidized silicon wafers and thinned down via a combination of grinding, polishing, and etching. Devices were then patterned with electron beam lithography using a ZEP520A ebeam resist (Zeon Corp.) along with an aluminum hard mask using Cl$_2$/BCl$_3$ and SF$_6$ dry etch gas chemistries. Structures were then first capped with flowable oxide (FOx-16, Dow Corning), followed by an additional SiO$_2$ layer through high-density plasma chemical vapor deposition (HDP-CVD), and were annealed at 800$^{\circ}$C in air. Facets for edge-coupling of devices were then created by wafersaw dicing. 

\section*{Acknowledgments}
This work is funded by the Defense Advanced Research Projects Agency under the LUMOS programme and by the IET A F Harvey Prize. J.Y. acknowledges support from the National Defense Science and Engineering Graduate (NDSEG) Fellowship. M.A.G. acknowledges the Albion Hewlett SGF and the NSF Graduate Research Fellowship. D.M.L. acknowledges the J. Hewes Crispin and Marjorie Holmes Crispin Stanford Graduate Fellowship (SGF) and the National Defense Science and Engineering Graduate (NDSEG) Fellowship. Part of this work was performed at the Stanford Nanofabrication Facility (SNF) and the Stanford Nano Shared Facilities (SNSF).


\bibliography{Reference}

\end{document}